\documentclass[11pt]{article}
\usepackage{graphicx}
\usepackage{amssymb}
\usepackage{amsmath, amsthm}
\newcommand{\ee}{\end{equation}}
\newcommand{\word}[1]{\,\,\mbox{#1}\,\,}
\newcommand{\reff}[1]{(\ref{#1})}
\newcommand{\beq}{\begin{equation}}
\newcommand{\eeq}[1]{\label{#1}\end{equation}}
\newcommand{\beg}{\begin{equation*}}
\newcommand{\eeg}{\end{equation*}}

\newcommand{\fivequad}{\qquad\qquad\qquad\qquad\qquad}
\newcommand{\eq}{\!=\!}
\newcommand{\p}{\!+\!}
\newcommand{\m}{\!-\!}

\newcommand{\sumprime}{\sideset{}{'}\sum}

\newcommand{\bsplit}{\begin{split}}
\newcommand{\esplit}{\end{split}}

\newcommand{\bino}[2]{\Bigl(\genfrac{}{}{0pt}{}{#1}{#2}\Bigr)}
\begin{document}
\def\theequation{\arabic{section}.\arabic{equation}}
\begin{titlepage}
\title{The perfect magnetic conductor (PMC) Casimir piston \\in $d\!+\!1$ dimensions}
\author{Ariel Edery\,\thanks{Email: aedery@ubishops.ca}\\
{\small\it Physics Department, Bishop's University}\\
{\small\it 2600 College Street, Sherbrooke, Qu\'{e}bec, Canada
J1M~0C8} and \\ \\ $^{1,2}$Valery Marachevsky\,\thanks{Email:
maraval@mail.ru}\\$^1${\small\it Laboratoire Kastler Brossel, CNRS,
ENS, UPMC,}\\{\small\it  Campus Jussieu case 74, 75252 Paris,
France}\\\\$^2${\small\it V. A Fock Institute of Physics, St.
Petersburg State University}\\{\small\it 198504 St. Petersburg,
Russia}}

\date{} \maketitle

\begin{abstract}
\noindent Perfect magnetic conductor (PMC) boundary conditions are
dual to the more familiar perfect electric conductor (PEC)
conditions and can be viewed as the electromagnetic analog of the
boundary conditions in the bag model for hadrons in QCD. Recent
advances and requirements in communication technologies have
attracted great interest in PMC's and Casimir experiments involving
structures that approximate PMC's may be carried out in the not too
distant future. In this paper, we make a study of the
zero-temperature PMC Casimir piston in $d\p 1$ dimensions. The PMC
Casimir energy is explicitly evaluated by summing over
$p+1$-dimensional Dirichlet energies where $p$ ranges from $2$ to
$d$ inclusively. We derive two exact $d$-dimensional expressions for
the Casimir force on the piston and find that the force is negative
(attractive) in all dimensions. Both expressions are applied to the
case of 2+1 and 3+1 dimensions. A spin-off from our work is a
contribution to the PEC literature: we obtain a useful alternative
expression for the PEC Casimir piston in 3+1 dimensions and also
evaluate the Casimir force per unit area on an infinite strip, a
geometry of experimental interest.
\end{abstract}
\setcounter{page}{1}
\end{titlepage}

\def\theequation{\arabic{section}.\arabic{equation}}

\section{Introduction}
\setcounter{page}{2}

Perfect magnetic conductor (PMC) boundary conditions are dual to the
more familiar perfect electric conductor (PEC) conditions. In 3+1
dimensions, the electric field ${\bf E}$ and the magnetic field
${\bf B}$ are zero inside a PEC and the condition at the surface is
${\bf n\cdot B}=0$ and ${\bf n \times E}=0$ where {\bf n} is the
vector normal to the surface. The conditions for PMC's is obtained
via the dual transformations ${\bf E\to H}$ and ${\bf B\to -D}$
where ${\bf H}$ is the magnetic field strength and ${\bf D}$ the
electric displacement. Inside a PMC, ${\bf D}$ and ${\bf H}$ are
zero, and the condition at the surface becomes ${\bf n\cdot D}=0$
and ${\bf n \times H}=0$. PMC boundary conditions can be generalized
to any dimension and are analogous to the boundary conditions in the
bag model for hadrons in QCD (see next section).

A distinctive property of a PMC is that its surface reflects
electromagnetic waves without phase change of the electric field, in
contrast to the $\pi$ phase change from a PEC \cite{antenna}. In the
last few years there has been a great interest in structures which
approximate PMC's because of their usefulness to communication
technologies, in particular low-profile antennas \cite{antenna}.
Casimir experiments involving structures which approximate PMC's
could therefore be carried out in the not too distant future. This
would be an exciting development.

In this paper, we make a study of the zero-temperature PMC Casimir
piston in a $d\p 1$ dimensional parallelepiped geometry. The PMC
Casimir energy can be expressed as sums over Dirichlet energies of
different dimensions and we perform this sum explicitly. We derive
two different and exact expressions for the $d$-dimensional Casimir
force on the piston. The second (alternative) expression is more
useful than the first expression when the plate separation is larger
than the dimension of the plates and vice-versa when the plate
separation is smaller. The PMC Casimir force is negative
(attractive) as in the PEC case (see refs.
\cite{Kardar}-\cite{Hertzberg}), Dirichlet case
\cite{Cavalcanti,Ariel2,Kardar,Hertzberg} and Neumann case
\cite{Kardar,Hertzberg,Ariel3}. The $d$-dimensional formulas (both
expressions) are applied to the case of 2+1 and 3+1 dimensions and
our results agree with previous Dirichlet and PEC results (see
section 3). A spin-off from our work on PMC's is a contribution to
the PEC literature. We obtain a novel alternative expression for the
3+1 dimensional PEC Casimir piston and also calculate the Casimir
force per unit area on an infinite strip, a case of experimental
interest.

The piston geometry has attracted considerable attention since the
original work of Cavalcanti \cite{Cavalcanti} because it resolves
the issue of surface divergences that often plague Casimir
calculations \cite{Weigel} and moreover, includes the non-trivial
effects of the exterior region. A Casimir piston contains an
interior and an exterior region and Cavalcanti showed explicitly for
the case of a massless scalar field in a 2+1 dimensional rectangular
cavity that the surface terms of the interior and exterior regions
canceled. He also showed that the Casimir force on the piston is
always negative regardless of the ratios of the two sides of the
rectangular region. This is in contrast to calculations that can
yield positive Casimir forces in rectangular geometries when no
exterior region is considered (see references in
\cite{Bordagreport}).

The PEC Casimir piston at zero-temperature in a 3+1-dimensional
square cavity was first studied in \cite{Kardar} and exact results
were obtained for the Casimir force on the piston. It was also shown
that the force was attractive. Expressions for arbitrary
cross-section valid for small plate separation were also derived.
Moreover, in that same work, the authors studied the Dirichlet and
Neumann piston in 3+1 dimensions obtaining results for small plate
separation (exact results were then obtained in
\cite{Ariel2,Hertzberg,Ariel3}; see paragraph below). The PEC work
was generalized further in \cite{Marachevsky1} (see also
\cite{Marachevsky2,Marachevsky3}) where exact results for arbitrary
cross sections at zero and finite temperature were first obtained
for arbitrary separation. The arbitrary cross-section results were
applied to both rectangular and circular cross-sections and explicit
expressions were obtained for these geometries. A positive feature
of the Casimir expressions in \cite{Marachevsky1} is that they are
manifestly negative. Using an optical path technique, finite
temperature results for the PEC rectangular piston with arbitrary
separation were later obtained in a different form
\cite{Hertzberg}(arbitrary cross section results valid for small
plate separation were also obtained). The physical meaning behind
the various terms that contribute to the Casimir energy of a
rectangular cavity was recently discussed as a three step process
involving piston interactions \cite{Marachevsky1A}. It is worth
noting that Casimir forces in a piston geometry can be repulsive
(positive) under various conditions. This was discussed in
\cite{Barton, Fulling}.

Besides the usual electromagnetic field, there is also good reason
to consider massless scalar fields in Casimir calculations. As
discussed in \cite{Kardar,Hertzberg}, the PEC Casimir energy can be
obtained from the Dirichlet and Neumann energy. Moreover, Casimir
results for massless scalar fields have been shown to have direct
application to physical systems such as Bose-Einstein condensates
\cite{Ariel4,Pomeau,Visser}. Higher-dimensional scalar field Casimir
calculations have also appeared in 6D supergravity theories
\cite{Burgess} and recently, the Casimir force on a piston with
extra compactified dimensions has been investigated \cite{Hongbo}.
As already mentioned above, scalar fields in the 3+1 dimensional
piston scenario were first studied in \cite{Kardar} and approximate
results valid for small plate separation were obtained. Exact
results for the zero-temperature 3+1 dimensional Dirichlet piston
with rectangular cross-section was then obtained in \cite{Ariel2}
via a multidimensional cut-off technique \cite{Ariel1}. Exact zero
and finite temperature results for the 3+1 Neumann and Dirichlet
piston with rectangular cross sections were then obtained in
\cite{Hertzberg}. Shortly thereafter, using a different technique,
two different expressions for the zero-temperature 3+1 Neumann
piston with rectangular cross section were obtained in
\cite{Ariel3}. The zero-temperature Dirichlet and Neumann Casimir
piston for parallelepiped geometries was solved exactly in arbitrary
dimensions in \cite{Ariel3}. The $d$-dimensional formulas were
applied to the 2+1 (and 3+1) dimensional Neumann piston bringing a
completion to Cavalcanti's original work in 2+1 dimensions
\cite{Cavalcanti}.
\section{PMC Casimir piston in d+1-dimensions}
\setcounter{equation}{0} The PMC boundary conditions ${\bf n \cdot
E}\eq0$ and ${\bf n\times B}\eq 0$ can be expressed as
$\eta^{\mu}F_{\mu\nu}=0$ where
$F_{\mu\nu}\equiv\partial_{\mu}A_{\nu}-\partial_{\nu}A_{\mu}$ is the
electromagnetic field tensor and  $\eta^{\mu}$ is a spacelike vector
normal to the surface. This is analogous to the boundary conditions
in the bag model for hadrons in QCD. We choose the gauge condition
$A_0\eq 0$ and $\partial^iA_i \eq 0$. The PMC condition
$\eta^{\mu}F_{\mu\nu}\eq 0$ together with the gauge condition
applies to any dimension. The mode decomposition for a
parallelepiped geometry in $d\p1$ dimensions with sides of lengths
$L_i$ where $i$ runs from $1$ to $d$ inclusively is given by
\cite{Wolfram}:\beq A_i=c_i \sin(k_i\,x_i)
\prod_{\substack{j=1\\j\ne i}}^d \cos(k_j\,x_j)\, e^{-i\omega
t}\eeq{mode_decomp} where $k_p=n_p\,\pi/L_p$, $n_p \ge 0 \in N$ and
$\omega\eq (k_p\,k^p)^{1/2}$ where $p$ runs from $1$ to $d$
inclusively. The PMC Casimir energy can be decomposed into sums of
Dirichlet energies of different dimensions. When all $n_i$'s are
non-zero, there are $d$ modes but the gauge condition reduces this
by $1$ yielding $d\m1$ independent modes. When one of the $n_i$'s is
zero, there are $d\m2$ independent modes. In general, there are $d\m
j$ independent modes when $j\m1$ $n_i$'s are zero (where $j$ runs
from $1$ to $d\m1$ inclusively). Each of those modes has the energy
of a scalar field in $d\m j\p 1$ dimensions obeying Dirichlet
boundary conditions. One must sum over all distinct sets of $d\m j\p
1$ lengths chosen among the $d$ lengths $L_1,L_2,..,L_d$. The
Casimir energy $E$ for PMC boundary conditions in a $d\p 1$
dimensional parallelepiped geometry with sides of length $L_1,
L_2,..,L_d$ can therefore be expressed as sums over Dirichlet(D)
Casimir energies $E_D$  (in units where $\hbar\!=\!c\!=\!1$)
\cite{Wolfram}: \beq E=\sum_{j=1}^{d-1}
\,(d-j)\,\,\xi^{d}_{\,i_1,.., i_{d-j+1}} \,E_{D_{\,i_1,..,
i_{d-j+1}}}\,.\eeq{PMC1} There is an implicit summation over the
integers $i_j$ in \reff{PMC1}. The ordered symbol
$\xi^{d}_{\,i_1,.., i_p}$, originally introduced in \cite{Ariel2},
is defined as \beq \xi^{\,d}_{\,i_1,.., i_p}=\begin{cases}
1&\word{if}
 i_1 \!<\!i_2\!<\! \ldots\! <i_p \,;\,1 \le i_p \le d\\ 0&
\word{otherwise}.
\end{cases}\eeq{order}
For $p=0$, $\xi^{d}_{\,i_1,.., i_p}$ is defined to be unity. The
ordered symbol ensures that the implicit sum over the $i_j$'s is
over all distinct sets $\{i_1,\ldots, i_p\}$, where the $i_j$'s are
integers that can run from $1$ to $d$ inclusively under the
constraint that $i_1\!<\!i_2\!<\!\cdots\!<i_p$. The superscript $d$
specifies the maximum value of $i_p$. For example, if $p=2$ and
$d=3$ then $\xi^{\,d}_{\,i_1,.., i_p}=\xi^{\,3}_{\,i_1, i_2}$ and
the non-zero terms are $\xi_{\,1,2}$ , $\xi_{\,1,3}$ and $
\xi_{\,2,3}$. This means the summation is over $\{i_1,i_2\}=(1,2),
(1,3)$ and $(2,3)$ so that
$\xi^{\,3}_{\,i_1,i_2}\,E_{i_1\,i_2}=E_{12}+E_{13}+E_{23}$. The
expression for the $d$-dimensional Dirichlet Casimir energy was
previously obtained and is given by \cite{Ariel2,Ariel3} \beq
E_{D_{1,2,..,d}}=\dfrac{\pi}{2^{d+1}}\sum_{p=0}^{d-1} (-1
)^{d+p}\,\,\xi^{\,d-1}_{\,k_1,.., k_p}\,\dfrac{L_{k_1}\ldots
L_{k_p}}{(L _d)^{p+1}}\big(Q_p
 +
R_{D_p}\big)\,\eeq{ED} where $Q_p$ is a function of $p$ and a
product of gamma and Riemann zeta functions: \beq Q_p=
\Gamma(\tfrac{p+2}{2})\,\pi^{\frac{-p-4}{2}}\, \zeta(p+2)\,.
\eeq{Qp} $R_{D_p}$ can be thought of as a remainder and is an
infinite sum over modified Bessel functions that converges rapidly
\beq R_{D_p}
=\sum_{n=1}^{\infty}\,\sumprime_{\substack{\ell_i=-\infty\\i=1,\ldots,
p}}^{\infty}\dfrac{2\,\,n^{\frac{p+1}{2}}}{\pi}\,\dfrac{\,K_{\frac{p+1}{2}}
\big(\,2\pi\,n\,\sqrt{(\ell_1\frac{L_{k_1}}{L_d})^2+\cdots+(\ell_p\,\frac{L_{k_p}}{L_d})^2}\,\,\,\big)}
{\left[(\ell_1\frac{L_{k_1}}{L_d})^2+\cdots+(\ell_p\frac{L_{k_p}}{L_d})^2\right]^{\tfrac{p+1}{4}}}\,.
\eeq{RD}

The prime on the sum in \reff{RD} means that the case when all
$\ell$'s are simultaneously zero
($\ell_1\!=\!\ell_2\!=\!\ldots\!=\!\ell_p\!=\!0$) is to be excluded.
There is an implicit summation over the $k_i$'s via the ordered
symbol $\xi_{\,k_1,.., k_p}$ defined in \reff{order}. Unlike $Q_p$,
$R_{D_p}$ does not depend only on $p$ but is also a function of the
ratios of lengths i.e.
$R_{D_p}=R_{D_p}(L_{k_1}/L_d,\ldots,L_{k_p}/L_d)$. Therefore, the
implicit summation over the $k_i$'s applies also to $R_{D_p}$. For
$p=0$, $R_{D_p}$ is defined to be zero and $\xi_{\,k_1,.., k_p}$ and
$L_{k_p}$ are defined to be unity so that $\xi^{\,d-1}_{\,k_1,..,
k_p}\,(L_{k_1}\ldots L_{k_p})/(L _d)^{p+1}=1/L_d$ for $p=0$.

The piston divides the volume into two regions: region I (inside)
and region II (outside). In region I, the $d$ sides are of length
$a_1,a_2,..,a_{d-1},a$ where $a$ is the plate separation. The $d$
sides are labeled in the following fashion: $L_1\!=\!a_1$,
$L_2\!=\!a_2$, $L_{d-1}\!=\!a_{d-1}$ and $L_d\!=\!a$. The Casimir
force depends on the derivative with respect to $a$ of the Casimir
energy and therefore only those terms which contain $L_d\!=\!a$ need
to be included. For the Casimir energy $E_{D_{\,i_1,.., i_{d-j+1}}}$
appearing in \reff{PMC1}, the length $L_d$ occurs when $i_{d-j+1}=d$
(recall that $i_1\!<\!i_2\!<\!..\!<\!i_{d-j+1}$). Therefore
\beq\xi^{d}_{\,i_1,.., i_{d-j+1}} \,E_{D_{\,i_1,..,
i_{d-j+1}}}=\xi^{d-1}_{\,i_1,.., i_{d-j}} \,E_{D_{\,i_1,..,
i_{d-j},d}}\,.\eeq{blank1} The formula for $E_{D_{\,i_1,..,
i_{d-j},d}}$ is obtained by replacing $d$ by $d\!-\!j\!+\!1$ and
$L_{k_1}$ by $L_{i_{k_1}}$, $L_{k_p}$ by $L_{i_{k_p}}$ and $L_d$ by
$L_{i_{d-j+1}}\!=\!L_d$ in \reff{ED}: \beq E_{D_{\,i_1,..,
i_{d-j},d}} =\dfrac{\pi}{2^{(d-j+2)}}\sum_{p=0}^{d-j} (-1
)^{d+p-j+1}\,\,\xi^{\,d-j}_{\,k_1,.., k_p}\,\dfrac{L_{i_{k_1}}\ldots
L_{i_{k_p}}}{(L _d)^{p+1}}\big( Q_p + R_{p}\big)\,\eeq{ED2} where
$R_p$ is equal to $R_{D_p}$ with $L_{k_1}$ replaced by
$L_{i_{k_1}}$, $L_{k_p}$ by $L_{i_{k_p}}$ and $L_d$ by
$L_{i_{d-j+1}}\!=\!L_d$.

To evaluate $\xi^{d-1}_{\,i_1,.., i_{d-j}} \,E_{D_{\,i_1,..,
i_{d-j},d}}$ we need to determine $\xi^{d-1}_{\,i_1,.., i_{d-j}}
\xi^{\,d-j}_{\,k_1,.., k_p}\,L_{i_{k_1}}\ldots L_{i_{k_p}}$. The
number of distinct sets $(i_1,.., i_{d-j})$ that can be generated by
$\xi^{d-1}_{\,i_1,.., i_{d-j}}$ is the binomial coefficient
$\bino{d-1}{d-j}$. The number of those sets that contain a
particular set $(i_{k_1},..,i_{k_p})$ is simply
$\bino{d-1-p}{d-j-p}$. We therefore obtain \beq \xi^{d-1}_{\,i_1,..,
i_{d-j}} \xi^{\,d-j}_{\,k_1,.., k_p}\,L_{i_{k_1}}\ldots L_{i_{k_p}}
=\bino{d\!-\!1\!-\!p}{d\!-\!j\!-\!p}\,\xi^{\,d-1}_{\,k_1,..,
k_p}\,L_{k_1}\ldots L_{k_p}\,.\eeq{binom} As an illustration
consider the case $d\!=\!5$, $p\!=\!2$ and $j\!=\!2$. Evaluating the
left hand side of \reff{binom} yields \beq
\begin{split}&\xi^{d-1}_{\,i_1,.., i_{d-j}} \xi^{\,d-j}_{\,k_1,..,
k_p}\,L_{i_{k_1}}\ldots
L_{i_{k_p}}=\xi^{4}_{\,i_1,i_2,i_3}\,\xi^{\,3}_{\,k_1,k_2}\,L_{i_{k_1}}
L_{i_{k_2}}=\xi^{4}_{\,i_1,i_2,i_3}(\xi^{\,3}_{1,2}L_{i_1}
L_{i_2}+\xi^{3}_{1,3}\,L_{i_1} L_{i_3}+\xi^{3}_{2,3}\,L_{i_2}
L_{i_3})\\&=\xi^{4}_{\,i_1,i_2,i_3}(L_{i_1} L_{i_2}+L_{i_1}
L_{i_3}+L_{i_2} L_{i_3})= \xi^{4}_{\,1,2,3}(L_{1} L_{2}+L_{1}
L_{3}+L_{2} L_{3})+\xi^{4}_{\,1,2,4}(L_{1} L_{2}+L_{1} L_{4}+L_{2}
L_{4})\\&+\xi^{4}_{\,1,3,4}(L_{1} L_{3}+L_{1} L_{4}+L_{3}
L_{4})+\xi^{4}_{\,2,3,4}(L_{2} L_{3}+L_{2} L_{4}+L_{3}
L_{4})\\&=2\,(L_{1} L_{2}+L_{1} L_{3}+L_{1} L_{4}+L_{2} L_{3}+L_{2}
L_{4}+L_{3} L_{4})\,.\end{split}\eeq{exampleL} The right hand side
of \reff{binom} yields \beq
\bino{d\!-\!1\!-\!p}{d\!-\!j\!-\!p}\,\xi^{\,d-1}_{\,k_1,..,
k_p}\,L_{k_1}\ldots L_{k_p} =
\bino{2}{1}\xi^{\,4}_{\,k_1,k_2}\,L_{k_1}L_{k_2} = 2\,(L_{1}
L_{2}+L_{1} L_{3}+L_{1} L_{4}+L_{2} L_{3}+L_{2} L_{4}+L_{3}
L_{4})\eeq{exampleR} which is equal to the result in
\reff{exampleL}.

The Casimir energy in region I, $E_I$, is obtained by substituting
\reff{blank1} into \reff{PMC1} and then using formula \reff{ED2} and
the equality \reff{binom}:  \beq\begin{split} E_I&=\sum_{j=1}^{d-1}
\sum_{p=0}^{d-j}(-1
)^{d+p-j+1}\dfrac{\pi}{2^{(d-j+2)}}(d-j)\bino{d\!-\!1\!-\!p}{d\!-\!j\!-\!p}\,\xi^{\,d-1}_{\,k_1,..,
k_p}\,\dfrac{L_{k_1}\ldots L_{k_p}}{(L_d)^{p+1}}\big( Q_p +
R_{p}\big)\\&=\sum_{p=0}^{d-1}\sum_{j=1}^{d-p} (-1
)^{d+p-j+1}\dfrac{\pi}{2^{(d-j+2)}}(d-j)\bino{d\!-\!1\!-\!p}{d\!-\!j\!-\!p}\,\xi^{\,d-1}_{\,k_1,..,
k_p}\,\dfrac{a_{k_1}\ldots a_{k_p}}{a^{p+1}}\big( Q_p +
R_{I_p}\big)\,.\end{split}\eeq{EI} Note that we have rearranged the
double sum, replaced $L_d$ by the plate separation $a$ and $L_{k_i}$
by $a_{k_i}$. The sum over $j$ can be readily evaluated and yields
\beq \sum_{j=1}^{d-p} (-1
)^{d+p-j+1}\dfrac{d-j}{2^{(d-j+2)}}\bino{d\!-\!1\!-\!p}{d\!-\!j\!-\!p}=
\dfrac{(d\!-\!1\!-\!2\,p)}{2^{d+1}}\,.\eeq{sumj}  We finally obtain
for region I \beq
E_I=\dfrac{\pi}{2^{d+1}}\sum_{p=0}^{d-1}\,(d\!-\!1\!-\!2\,p)\,\,\xi^{\,d-1}_{\,k_1,..,
k_p}\,\dfrac{a_{k_1}\ldots a_{k_p}}{a^{p+1}}\,\big( Q_p +
R_{I_p}\big)\eeq{EIP} where $Q_p$ is given by \reff{Qp} and
$R_{I_p}$ is obtained from $R_{D_p}$ by replacing $L_{k_i}$ by
$a_{k_i}$ and $L_d$ by the plate separation $a$ i.e.  \beq R_{I_p}
=\sum_{n=1}^{\infty}\,\sumprime_{\substack{\ell_i=-\infty\\i=1,\ldots,
p}}^{\infty}\dfrac{2\,\,n^{\frac{p+1}{2}}}{\pi}\,\dfrac{\,K_{\frac{p+1}{2}}
\big(\,2\pi\,n\,\sqrt{(\ell_1\frac{a_{k_1}}{a})^2+\cdots+(\ell_p\,\frac{a_{k_p}}{a})^2}\,\,\,\big)}
{\left[(\ell_1\frac{a_{k_1}}{a})^2+\cdots+(\ell_p\frac{a_{k_p}}{a})^2\right]^{\tfrac{p+1}{4}}}\,.
\eeq{RIP}

 We now evaluate the PMC Casimir energy in region II, $E_{II}$. In region
 II, the $d$ lengths are $s-a$, $a_1,a_2,..,a_{d-1}$ (i.e. the same
 lengths as in region I except that $a$ is replaced by $s\!-\!a$).
 We label the lengths in region II as
 $L_1=s\!-\!a$, $L_2=a_1$, $L_3=a_2$ and
 $L_d =a_{d-1}$. The Casimir force is obtained by taking the
 derivative with respect to $a$ so that only terms that contain
 $L_1$ need to be included in the Casimir energy. We therefore set
 $i_1=1$ in \reff{PMC1} which yields
\beq E_{I\!I}=\sum_{j=1}^{d-1} \,(d-j)\,\,\xi^{d}_{\,1,i_2,i_3,..,
i_{d-j+1}} \,E_{D_{\,1,i_2,i_3,.., i_{d-j+1}}}\,\eeq{EPMCII} where
the Dirichlet expression is given by \reff{ED}: \beq
E_{D_{\,1,i_2,i_3.., i_{d-j+1}}}
=\dfrac{\pi}{2^{(d-j+2)}}\sum_{p=1}^{d-j} (-1
)^{d+p-j+1}\,\,\xi^{\,d-j}_{\,1,k_2,k_3,..,
k_p}\,\dfrac{L_1\,L_{i_{k_2}}\ldots L_{i_{k_p}}}
{(L_{i_{d-j+1}})^{p+1}}\big( Q_p + R_{{I\!I}_p}\big)\,. \eeq{ED33}
$R_{{II}_p}$ is obtained from $R_{D_p}$ eq. \reff{RD} with $L_{k_1}$
replaced by $L_1\!=\!s\!-\!a$, $L_{k_p}$ by $L_{i_{k_p}}$ and $L_d$
by $L_{i_{d-j+1}}$. Substituting \reff{ED33} into \reff{EPMCII}
yields: \beq E_{I\!I}=\sum_{j=1}^{d-1}\,\sum_{p=1}^{d-j} (-1
)^{d+p-j+1}\,\dfrac{\pi}{2^{(d-j+2)}}\,(d-j)\,\xi^{d}_{\,1,i_2,i_3,..,
i_{d-j+1}}\,\xi^{\,d-j}_{\,1,k_2,k_3,..,
k_p}\,\dfrac{L_1\,L_{i_{k_2}}\ldots L_{i_{k_p}}}
{(L_{i_{d-j+1}})^{p+1}}\big( Q_p + R_{{I\!I}_p}\big)\eeq{j2} In
$\xi^{d}_{\,1,i_2,i_3,.., i_{d-j+1}}$ the value of $i_{d-j+1}$
ranges from $d\!-\!j\!+\!1$ to $d$ inclusively. We can therefore
replace $i_{d-j+1}$ with $d-q$ and sum $q$ from $0$ to $j-1$
yielding \beq\begin{split} \xi^{d}_{\,1,i_2,i_3,..,
i_{d-j+1}}\,\xi^{\,d-j}_{\,1,k_2,k_3,..,
k_p}\,\dfrac{L_1\,L_{i_{k_2}}\ldots L_{i_{k_p}}}{(L
_{i_{d-j+1}})^{p+1}}&= \sum_{q=0}^{j-1} \xi^{d-q-1}_{\,1,i_2,i_3,..,
i_{d-j},d-q}\,\xi^{\,d-j}_{\,1,k_2,k_3,..,
k_p}\,\dfrac{L_1\,L_{i_{k_2}}\ldots L_{i_{k_p}}}{(L
_{d-q})^{p+1}}\\&=\sum_{q=0}^{j-1}
\bino{d\!-\!q\!-\!1\!-\!p}{d\!-\!j\!-\!p}\,\xi^{\,d-q-1}_{\,1,k_2,k_3,..,
k_p}\,\dfrac{L_1\,L_{k_2}\ldots L_{k_p}}{(L_{d-q})^{p+1}}\,
\end{split}\eeq{vs}
where the binomial coefficient follows from the same reasoning given
above \reff{binom}.

Substituting the above into \reff{j2} yields \beq\begin{split}
E_{I\!I}\!&=\!\sum_{j=1}^{d-1}\sum_{p=1}^{d-j} \sum_{q=0}^{j-1}(-1
)^{d+p-j+1}\,\dfrac{\pi\,(d-j)}{2^{(d-j+2)}}\,\bino{d\!-\!q\!-\!1\!-\!p}{d\!-\!j\!-\!p}\,\xi^{\,d-q-1}_{\,1,k_2,k_3,..,
k_p}\,\dfrac{L_1\,L_{k_2}\ldots L_{k_p}} {(L_{d-q})^{p+1}}\big( Q_p
+ R_{{I\!I}_p}\big)\\&=
\sum_{p=1}^{d-1}\sum_{q=0}^{d-p-1}\sum_{j=q+1}^{d-p} (-1
)^{d+p-j+1}\,\dfrac{\pi\,(d-j)}{2^{(d-j+2)}}\,\bino{d\!-\!q\!-\!1\!-\!p}{d\!-\!j\!-\!p}\,\xi^{\,d-q-1}_{\,1,k_2,k_3,..,
k_p}\,\dfrac{(s-a)\,a_{k_2-1}\ldots a_{k_p-1}}
{(a_{d-q-1})^{p+1}}\big( Q_p +
R_{{I\!I}_p}\big)\,\end{split}\eeq{j3} where the triple sum has been
rearranged into an equivalent form and the lengths corresponding to
the $L_i$'s was substituted. The sum over $j$ yields \beq
\sum_{j=q+1}^{d-p}(-1
)^{d+p-j+1}\,\dfrac{(d-j)}{2^{(d-j+2)}}\,\bino{d\!-\!q\!-\!1\!-\!p}{d\!-\!j\!-\!p}\,
= \dfrac{(d\!-\!1\!-\!2\,p\!-\!q)}{2^{d-q+1}}\,\eeq{sumj2} and we
finally obtain the PMC Casimir energy $E_{I\!I}$ for region II:\beq
E_{I\!I}=\sum_{p=1}^{d-1}\,\,\sum_{q=0}^{d-p-1}
\dfrac{\pi\,(d\!-\!1\!-\!2p\!-\!q)}{2^{d-q+1}}\,\,\xi^{\,d-q-1}_{\,1,k_2,k_3,..,
k_p}\,\dfrac{(s-a)\,a_{k_2-1}\ldots a_{k_p-1}}
{(a_{d-q-1})^{p+1}}\big( Q_p + R_{{I\!I}_p}\big)\,\eeq{EIIPMC} where
$Q_p$ is again given by \reff{Qp} and $R_{I\!I_p}$ is obtained from
eq.\reff{RD} with $L_{k_1}$ replaced by $L_1\!=\!s\!-\!a$, $L_{k_p}$
by $a_{k_p-1}$ and $L_d$ by $a_{d-q-1}$ i.e.\beq R_{I\!I_p}
=\sum_{n=1}^{\infty}\,\sumprime_{\substack{\ell_i=-\infty\\i=1,\ldots,
p}}^{\infty}\dfrac{2\,\,n^{\frac{p+1}{2}}}{\pi}\,\dfrac{\,K_{\frac{p+1}{2}}
\big(\,2\pi\,n\,\sqrt{(\ell_1\frac{s-a}{a_{d-q-1}})^2+\cdots+(\ell_p\,\frac{a_{k_p-1}}{a_{d-q-1}})^2}\,\,\,\big)}
{\left[(\ell_1\frac{s-a}{a_{d-q-1}})^2+\cdots+(\ell_p\frac{a_{k_p-1}}{a_{d-q-1}})^2\right]^{\tfrac{p+1}{4}}}\,.
\eeq{RIIp}

\subsection{PMC Casimir force expressions in d+1 dimensions}
The Casimir force is obtained by taking the negative derivative with
respect to the plate separation $a$ of the Casimir energy. In region
I the Casimir energy is given by \reff{EIP} together with \reff{Qp}
and \reff{RIP}. The Casimir force in region I, $F_I$ is given by
\beq\begin{split} F_I&=-\dfrac{\partial\,E_I}{\partial\,a}\\&=
\dfrac{\pi}{2^{d+1}}\sum_{p=0}^{d-1}\,(d\!-\!1\!-\!2p)\,(p+1)\,\,\xi^{\,d-1}_{\,k_1,..,
k_p}\,\dfrac{a_{k_1}\ldots
a_{k_p}}{a^{p+2}}\,\,\Gamma(\tfrac{p+2}{2})\,\,\pi^{\frac{-p-4}{2}}\,\,
\zeta(p+2)\, -
\,\dfrac{\partial\,R_I}{\partial\,a}\end{split}\eeq{FI} where \beq
\begin{split}\dfrac{\partial\,R_I}{\partial\,a}&=
\dfrac{\partial}{\partial\,a}\Bigg\{\dfrac{\pi}{2^{d+1}}\sum_{p=1}^{d-1}\,(d\!-\!1\!-\!2p)\,\,\xi^{\,d-1}_{\,k_1,..,
k_p}\dfrac{\,a_{k_1}\ldots a_{k_p}}{a^{p+1}}\,
R_{I_p}\Bigg\}\\&=\dfrac{\pi}{2^{d-1}}\sum_{p=1}^{d-1}\,\sum_{n=1}^{\infty}\,\sumprime_{\substack{\ell_i=-\infty\\i=1,\ldots,
p}}^{\infty}(d\!-\!1\!-\!2p)\,\,\xi^{\,d-1}_{\,k_1,..,
k_p}\,a_{k_1}\ldots
a_{k_p}\,n^{\frac{p+3}{2}}\,\dfrac{\,K_{\frac{p-1}{2}}
\big(\,\frac{2\pi\,n}{a}\,\sqrt{(\ell_1\,a_{k_1})^2+\cdots+(\ell_p\,a_{k_p})^2}\,\,\,\big)}
{a^{\frac{p+5}{2}}\,\left[(\ell_1\,a_{k_1})^2+\cdots+(\ell_p\,a_{k_p})^2\right]^{\tfrac{p-1}{4}}}\,.
\end{split}\eeq{RI}
The Casimir energy in region II is given by \reff{EIIPMC} together
with \reff{Qp} and \reff{RIIp}. We are interested in the case when
the outside region $II$ is infinite i.e. $s\to \infty$. The Casimir
force in region $II$ is then given by\beq\begin{split}
F_{I\!I}&=-\lim_{s\to\infty}\dfrac{\partial\,E_{I\!I}}{\partial\,a}\\&=
\sum_{p=1}^{d-1}\,\,\sum_{q=0}^{d-p-1}
\dfrac{\pi\,(d\!-\!1\!-\!2p\!-\!q)}{2^{d-q+1}}\,\,\xi^{\,d-q-1}_{\,1,k_2,k_3,..,
k_p}\,\dfrac{a_{k_2-1}\ldots a_{k_p-1}}
{(a_{d-q-1})^{p+1}}\\&\qquad\qquad\qquad\qquad\qquad\qquad\Big\{
\Gamma(\tfrac{p+2}{2})\,\,\pi^{\frac{-p-4}{2}}\,\,
\zeta(p+2)\,-\lim_{s\to\infty}\,\dfrac{\partial\,}{\partial\,a}\big\{(s-a)\,R_{{I\!I}_p}\big\}\Big\}\\&=\sum_{p=1}^{d-1}\,\,\sum_{q=0}^{d-p-1}
\dfrac{\pi\,(d\!-\!1\!-\!2p\!-\!q)}{2^{d-q+1}}\,\,\xi^{\,d-q-1}_{\,1,k_2,k_3,..,
k_p}\,\dfrac{a_{k_2-1}\ldots a_{k_p-1}} {(a_{d-q-1})^{p+1}}\,\Big\{
\Gamma(\tfrac{p+2}{2})\,\,\pi^{\frac{-p-4}{2}}\,\,
\zeta(p+2)\,+R_{{I\!I}_p}(\ell_1\!=\!0)\Big\}\end{split}\eeq{FII}
where $R_{{I\!I}_p}(\ell_1\!=\!0)$ means $R_{{I\!I}_p}$ evaluated
with $\ell_1\!=\!0$: \beq R_{I\!I_p}(\ell_1\!=\!0)
=\sum_{n=1}^{\infty}\,\sumprime_{\substack{\ell_i=-\infty\\i=2,\ldots,
p}}^{\infty}\dfrac{2\,\,n^{\frac{p+1}{2}}}{\pi}\,\dfrac{\,K_{\frac{p+1}{2}}
\big(\,2\pi\,n\,\sqrt{(\ell_2\frac{a_{k_2-1}}{a_{d-q-1}})^2+\cdots+(\ell_p\,\frac{a_{k_p-1}}{a_{d-q-1}})^2}\,\,\,\big)}
{\left[(\ell_2\frac{a_{k_2-1}}{a_{d-q-1}})^2+\cdots+(\ell_p\frac{a_{k_p-1}}{a_{d-q-1}})^2\right]^{\tfrac{p+1}{4}}}\,.
\eeq{RIIp2}

The Casimir force $F_{\!P\!M\!C}$ on the piston with perfect
magnetic conductor boundary conditions is finally obtained by adding
$F_I$ and $F_{I\!I}$:\beq\begin{split} F_{\!P\!M\!C}&=
\dfrac{\pi}{2^{d+1}}\sum_{p=0}^{d-1}\,(d\!-\!1\!-\!2p)\,(p+1)\,\,\xi^{\,d-1}_{\,k_1,..,
k_p}\,\dfrac{a_{k_1}\ldots
a_{k_p}}{a^{p+2}}\,\,\Gamma(\tfrac{p+2}{2})\,\,\pi^{\frac{-p-4}{2}}\,\,
\zeta(p+2)\, - \,\dfrac{\partial\,R_I}{\partial\,a}
\\&\qquad\qquad+\sum_{p=1}^{d-1}\,\,\sum_{q=0}^{d-p-1}
\dfrac{\pi\,(d\!-\!1\!-\!2p\!-\!q)}{2^{d-q+1}}\,\,\xi^{\,d-q-1}_{\,1,k_2,k_3,..,
k_p}\,\dfrac{a_{k_2-1}\ldots a_{k_p-1}}
{(a_{d-q-1})^{p+1}}\,\\&\qquad\qquad\qquad\qquad\qquad\qquad\qquad\qquad\qquad\Big\{
\Gamma(\tfrac{p+2}{2})\,\,\pi^{\frac{-p-4}{2}}\,\,
\zeta(p+2)\,+R_{{I\!I}_p}(\ell_1\!=\!0)\Big\}\,\end{split}\eeq{FPMC}
where $\partial\,R_I/\partial\,a$ is given by \reff{RI} and
$R_{{I\!I}_p}(\ell_1\!=\!0)$ by \reff{RIIp2}. The PMC Casimir force
in any spatial dimension $d$ and for arbitrary lengths of the sides
of the parallelepiped can be obtained via equation \reff{FPMC}
together with \reff{RI} and \reff{RIIp2}. The force is automatically
negative (attractive) because it is obtained from a sum over
Dirichlet Casimir piston forces which are negative.

We now discuss the rate of convergence of \reff{FPMC}. Note that
\reff{FPMC} contains two different kinds of terms: a finite sum over
analytical terms and infinite sums over modified Bessel functions.
The analytical terms contain inverse powers of the plate separation
$a$ (i.e. $1/a^{p+2}$) multiplied by gamma and Riemann zeta
functions. A finite sum of those terms is trivial to evaluate and
there are no convergence issues. The next term, $\partial
R_I/\partial a$ given by \reff{RI}, contains infinite sums over
modified Bessel functions. The ratios of lengths in the argument of
the modified Bessel functions have the plate separation $a$ in the
denominator. If $a$ is the smallest length, the modified Bessel
functions are tiny and the sum converges exponentially fast (only a
few terms need to be summed in \reff{RI} to reach convergence).
However, if the plate separation $a$ is the largest length (e.g.
square plates with sides of 1 micron separated by 10 microns), the
modified Bessel functions can be large and converge slowly. In the
large $a$ limit where $a_{k_i}/a<\!<1$, a large number of terms
would need to be summed in \reff{RI} to achieve convergence. Simply
put, when $a$ is large, it is not computationally efficient to use
\reff{FPMC} to evaluate the Casimir force.

By using the invariance of the Casimir energy under permutation of
lengths, it is possible to derive an alternative expression for the
PMC Casimir force  $F_{P\!M\!C}^{alt}$ that yields the same force as
\reff{FPMC} but converges exponentially fast when the plate
separation $a$ is the largest length. This expression is derived in
appendix A and is given by \reff{FPMCalt} together with
\reff{RIpalt}. In \reff{RIpalt}, the plate separation $a$ appears in
the ${\it numerator}$ in the argument of the modified Bessel
functions so that the infinite sums converge exponentially fast when
$a$ is the largest length. Computationally it is better to use the
alternative expression \reff{FPMCalt} instead of the above
expression \reff{FPMC} to calculate the PMC Casimir force when the
plate separation $a$ is the largest length and vice versa if $a$ is
the smallest length. The main results of this paper are the two
different expressions for the PMC Casimir force on the piston:
equations \reff{FPMC} and \reff{FPMCalt}.

\section{Applications: the 2+1 and 3+1 dimensional PMC Casimir piston}

As an illustration of how to apply the $d\!+\!1$-dimensional PMC
Casimir formula \reff{FPMC} or the alternative expression
\reff{FPMCalt} we consider 2+1 and 3+1 dimensions. The case of 2+1
dimensions is the simplest non-trivial case where equations
\reff{FPMC} and \reff{FPMCalt} can be applied. From \reff{PMC1}, we
see that in two spatial dimensions, the PMC Casimir energy is
equivalent to the Dirichlet energy. In three spatial dimensions (and
only in three), the PMC Casimir energy is equal to the PEC Casimir
energy. This can be seen most transparently in the transverse
electric (TE) and transverse magnetic (TM) decomposition that exists
in 3+1 dimensions. The PEC Casimir energy in 3+1 is half the sum
over all modes of the eigenfrequencies $\omega_{TE}$ and
$\omega_{TM}$ (see \cite{Marachevsky1} for a recent application of
the TE/TM decomposition in a piston geometry of arbitrary cross
section). The eigenfrequencies in the PMC case are obtained by
simply switching $\omega_{TE}$ for $\omega_{TM}$ and vice-versa
leaving the sum $\omega_{TE}\p\omega_{TM}$ unchanged. A strong
confirmation of our $d$-dimensional technique and PMC formulas is
that our 2+1 and 3+1 dimensional results are in agreement with
previous Dirichlet and PEC results respectively. An important
spin-off from our work is that we obtain an alternative expression
for the PEC Casimir piston in 3+1 dimensions and also obtain the
Casimir force per unit area for the special case of an infinite
strip.

\subsection{2+1 dimensions}

In $2\!+\!1$ dimensions we use $d=2$ in \reff{FPMC}. The two lengths
are $a_1=b$ and the plate separation $a$.  We evaluate the three
terms in \reff{FPMC} separately. The first term is \beq\begin{split}
\dfrac{\pi}{8}\sum_{p=0}^{1}\,(1-2p)(p+1)\xi^{\,1}_{\,k_1,..,
k_p}\,\dfrac{a_{k_1}\ldots
a_{k_p}}{a^{p+2}}\,\,\Gamma(\tfrac{p+2}{2})\,\,\pi^{\frac{-p-4}{2}}\,\,
\zeta(p+2)= \dfrac{\pi}{48\,a^2}- \dfrac{\zeta(3)\,b}{8\pi
a^3}\,.\end{split}\eeq{F1} The second term is evaluated via
eq.\reff{RI} with $d=2$: \beq- \,\dfrac{\partial\,R_I}{\partial\,a}=
\dfrac{\pi\,b}{a^3} \,\sum_{n=1}^{\infty}\,\sum_{\ell=1}^{\infty}\,
n^{2}\,K_{0} \Big(\,\frac{2\pi\,n\,\ell\,b}{a}\Big) \,.\eeq{F2} The
third term yields (only the $p\!=\!1$, $q\!=\!0$ case needs to be
evaluated): \beq
\begin{split}
&\dfrac{\pi\,(-1)}{2^{3}}\,\,\dfrac{1} {(a_{1})^{p+1}}\,\Big\{
\Gamma(\tfrac{3}{2})\,\,\pi^{\frac{-5}{2}}\,\,
\zeta(3)\,+R_{{I\!I}_p}(\ell_1\!=\!0)\Big\}\\&=
-\dfrac{\zeta(3)}{16\pi \,b^2}\,\end{split}\eeq{F3} where
$R_{{I\!I}_p}(\ell_1\!=\!0)$ is zero for $p=1$ (it starts at $p=2$).
The PMC Casimir force on the piston in $2+1$ dimensions is given by
summing all three terms:

\beq F_{\!P\!M\!C}=- \dfrac{\zeta(3)\,b}{8\pi
a^3}+\dfrac{\pi}{48\,a^2}+ \dfrac{\pi\,b}{a^3}
\,\sum_{n=1}^{\infty}\,\sum_{\ell=1}^{\infty}\, n^{2}\,K_{0}
\Big(\,\frac{2\pi\,n\,\ell\,b}{a}\Big)-\dfrac{\zeta(3)}{16\pi
\,b^2}\,.\eeq{FPMC2D} In the limit of infinite parallel lines, i.e.
$b\!\to\!\infty$, the force per unit length tends to $-\zeta(3)/8\pi
a^3$.

We now calculate the Casimir force using the alternative expression
\reff{FPMCalt} together with \reff{RIpalt}. For $d=2$,  we only need
to evaluate the term $(p\!=\!1,q\!=\!0)$ in \reff{FPMCalt}: \beq
F_{PMC}^{alt}=\dfrac{\pi}{8\,b^2}\dfrac{\partial\,}{\partial\,a}\Big\{a\,\sum_{n=1}^{\infty}\sum_{\ell=1}
^{\infty} \dfrac{4}{\pi}\dfrac{n\,b}{\ell\,a}
K_1\big(\frac{2\,\pi\,n\,\ell\,a}{b}\big)\Big\}=\dfrac{1}{2\,b}\sum_{n=1}^{\infty}\sum_{\ell=1}^{\infty}
\dfrac{n}{\ell}\,\dfrac{\partial\,}{\partial\,a}K_1\big(\frac{2\,\pi\,n\,\ell\,a}{b}\big)\,.
\eeq{Falt2D}

Eqs. \reff{FPMC2D} and \reff{Falt2D} are both in agreement with
those obtained by Cavalcanti \cite{Cavalcanti} for Dirichlet
boundary conditions in $2\!+\!1$ dimensions and this provides an
independent confirmation of our general PMC formulas \reff{FPMC} and
\reff{FPMCalt}.

\subsection{3+1 dimensions}

In $3+1$ dimensions we set $d=3$ in \reff{FPMC}. The three lengths
are $a_1=c$, $a_2=b$ and the plate separation $a$. Again we evaluate
the three terms in \reff{FPMC} separately. The first term yields
\beq\begin{split}
\dfrac{\pi}{16}\sum_{p=0}^{2}\,(2-2p)(p+1)\xi^{\,2}_{\,k_1,..,
k_p}\,\dfrac{a_{k_1}\ldots
a_{k_p}}{a^{p+2}}\,\,\Gamma(\tfrac{p+2}{2})\,\,\pi^{\frac{-p-4}{2}}\,\,
\zeta(p+2)= \dfrac{\pi}{48\,a^2}- \dfrac{3\,b\,c\,\zeta(4)}{8\pi^2
a^4}\,.\end{split}\eeq{F11} The second term is given by \beq
\begin{split}-\dfrac{\partial\,R_I}{\partial\,a}&=-\dfrac{\pi}{4}\sum_{p=1}^{2}\,\sum_{n=1}^{\infty}\,
\sumprime_{\substack{\ell_i=-\infty\\i=1,\ldots,
p}}^{\infty}(2-\!2p)\,\,\xi^{\,2}_{\,k_1,.., k_p}\,a_{k_1}\ldots
a_{k_p}\,n^{\frac{p+3}{2}}\,\dfrac{\,K_{\frac{p-1}{2}}
\big(\,\frac{2\pi\,n}{a}\,\sqrt{(\ell_1\,a_{k_1})^2+\cdots+(\ell_p\,a_{k_p})^2}\,\,\,\big)}
{a^{\frac{p+5}{2}}\,\left[(\ell_1\,a_{k_1})^2+\cdots+(\ell_p\,a_{k_p})^2\right]^{\tfrac{p-1}{4}}}\,
\\&=\dfrac{\pi\,b\,c}{2\,a^{7/2}}\,\sum_{n=1}^{\infty}\,\sumprime_{\substack{\ell_1,\ell_2=-\infty\\}}^{\infty}
n^{\frac{5}{2}}\,\dfrac{\,K_{\frac{1}{2}}
\big(\,\frac{2\pi\,n}{a}\,\sqrt{(\ell_1\,c)^2+(\ell_2\,b)^2}\,\big)}
{\,\left[(\ell_1\,c)^2+(\ell_2\,b)^2\right]^{\tfrac{1}{4}}}
\end{split}\eeq{RI1}
and the third term yields \beq\begin{split}
&\sum_{p=1}^{2}\,\,\sum_{q=0}^{2-p}
\dfrac{\pi\,(2-\!2p\!-\!q)}{2^{4-q}}\,\,\xi^{\,2-q}_{\,1,k_2,k_3,..,
k_p}\,\dfrac{a_{k_2-1}\ldots a_{k_p-1}}
{(a_{{}_{2\!-q}})^{p+1}}\Big\{
\Gamma(\tfrac{p+2}{2})\,\,\pi^{\frac{-p-4}{2}}\,\,
\zeta(p+2)\,+R_{{I\!I}_p}(\ell_1\!=\!0)\Big\}\\&= -
\dfrac{\zeta(3)}{16\,\pi c^2}- \dfrac{\zeta(4)\,c}{8\pi^2 b^3} -
\dfrac{1}{2\,c^{1/2}}
\,\sum_{n=1}^{\infty}\,\sum_{\ell=1}^{\infty}\,
\Big(\dfrac{n}{\ell\,b}\Big)^{3/2}\,K_{3/2}
\Big(\,\frac{2\pi\,n\,\ell\,c}{b}\Big)\,.
\end{split}\eeq{F33}
The PMC Casimir force in $3+1$ dimensions is obtained by summing all
three terms i.e.\beq\begin{split}
F_{\!P\!M\!C}&=\dfrac{\pi}{48\,a^2}-
\dfrac{3\,\zeta(4)\,b\,c}{8\pi^2
a^4}+\dfrac{\pi\,b\,c}{2\,a^{7/2}}\,\sum_{n=1}^{\infty}\,\sumprime_{\substack{\ell_1,\ell_2=-\infty\\}}^{\infty}
n^{\frac{5}{2}}\,\dfrac{\,K_{\frac{1}{2}}
\big(\,\frac{2\pi\,n}{a}\,\sqrt{(\ell_1\,c)^2+(\ell_2\,b)^2}\,\big)}
{\,\left[(\ell_1\,c)^2+(\ell_2\,b)^2\right]^{\tfrac{1}{4}}}\\&-
\dfrac{\zeta(3)}{16\,\pi c^2}- \dfrac{\zeta(4)\,c}{8\pi^2 b^3} -
\dfrac{1}{2\,c^{1/2}}
\,\sum_{n=1}^{\infty}\,\sum_{\ell=1}^{\infty}\,
\Big(\dfrac{n}{\ell\,b}\Big)^{3/2}\,K_{3/2}
\Big(\,\frac{2\pi\,n\,\ell\,c}{b}\Big)\,.\end{split}\eeq{finito}

Though expressed in a different form, equation \reff{finito} is in
numerical agreement with previous PEC results in 3+1 dimensions
refs.\cite{Kardar}-\cite{Hertzberg}. This provides another
independent confirmation of our $d$-dimensional equations.

To obtain the alternative expression for the Casimir force, we
substitute $d=3$ in \reff{FPMCalt}: \beq\begin{split}
 F_{\!P\!M\!C}^{alt}&=-\sum_{p=1}^{2}\,\,\sum_{q=0}^{2-p}
\dfrac{\pi\,(2\!-\!2p\!-\!q)}{2^{4-q}}\,\,\xi^{\,2-q}_{\,1,k_2,k_3,..,
k_p}\,\dfrac{a_{k_2-1}\ldots a_{k_p-1}}
{(a_{2-q})^{p+1}}\dfrac{\partial\,}{\partial\,a}\big\{a\,R_{I_p}^{alt}(\ell_1\!\ne\!0)\big\}
\\&=\dfrac{1}{2\,c}\,\sum_{n=1}^{\infty}\sum_{\ell=1}^{\infty}\dfrac{n}{\ell}
\dfrac{\partial\,}{\partial\,a}K_1\big(\frac{2\,\pi\,n\,\ell\,a}{c}\big)\\&\qquad\qquad
+\,\dfrac{\partial\,}{\partial\,a}
\Bigg\{\dfrac{a\,c}{2}\sum_{n=1}^{\infty}\,\sum_{\ell_1=1}^{\infty}\sum_{\substack{\ell_2=-\infty\\}}^{\infty}
\Big(\dfrac{n}{b}\Big)^{3/2}\,\dfrac{\,K_{\frac{3}{2}}
\big(\,\frac{2\pi\,n}{b}\,\sqrt{(\ell_1\,a)^2+ (\ell_2\,c)^2}
\,\big)}{\left[(\ell_1\,a)^2
+(\ell_2\,c)^2\right]^{\tfrac{3}{4}}}\Bigg\}\,.
\end{split}
\eeq{FPMCalt3} The alternative expression \reff{FPMCalt3} yields the
same value as the original expression \reff{finito} but converges
much faster if $a$ is larger than $b$ and $c$. Note that
\reff{FPMCalt3} is also an alternative expression for the PEC
Casimir piston in 3+1 dimensions. A spin-off from our work is
therefore a novel expression for the PEC piston that is highly
useful (converges fast) when $a$ is larger than $b$ and $c$.

\subsubsection{Infinite strip}
\begin{figure}[ht]
\begin{center}
\includegraphics[scale=0.80]{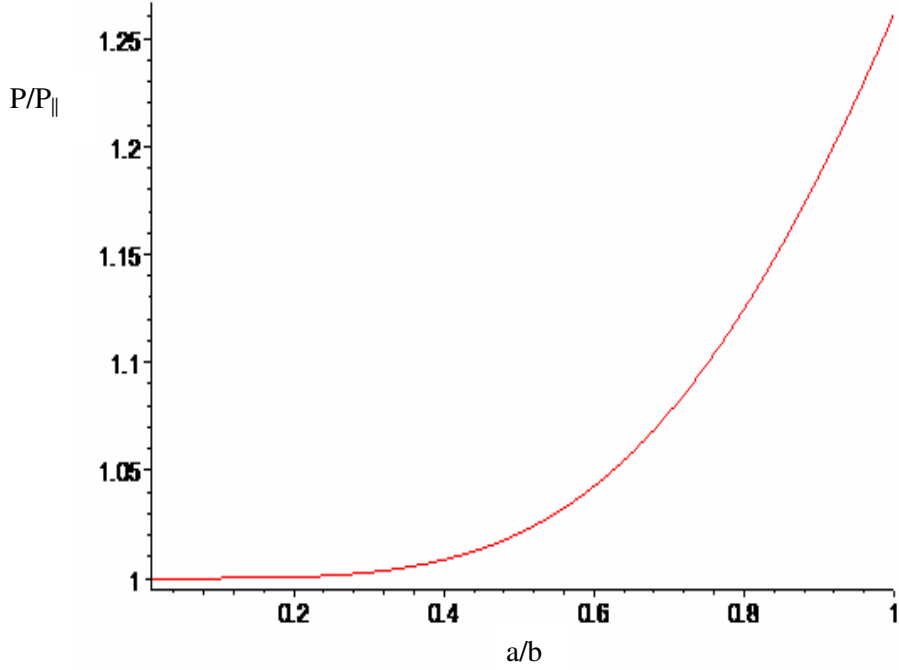}
\caption{Casimir pressure on infinite strip versus $a/b$ (in units
of $P_{||}$).}
\end{center}
\end{figure}
In this section we consider the special case of an infinite strip
where one side of the plates is of finite length and the other side
is infinitely long (yielding translation invariance along that
direction). An accurate measurement of the Casimir force between
parallel metallic surfaces was performed only a few years ago
\cite{Onofrio}. The infinite strip, being closely related in
geometry, should therefore be of experimental interest. The 3+1
dimensional Casimir force given by eq. \reff{finito} is invariant
under exchange of the two sides $b$ and $c$ and without loss of
generality we take $b$ to be finite and let $c\to\infty$. In this
limit, the term containing $K_{3/2}$ in \reff{finito} is zero and
the term containing $K_{1/2}$ is zero except when $\ell_1$ equals
zero. This yields a Casimir force per unit area (or pressure) of
\beq P\equiv\lim_{c\to\infty} \,\dfrac{F}{b\,c}=-
\dfrac{3\,\zeta(4)}{8\pi^2 a^4}- \dfrac{\zeta(4)}{8\pi^2
b^4}+\dfrac{\pi}{a^{7/2}}\,\sum_{n=1}^{\infty}\,\sum_{\ell=1}^{\infty}\,\,
n^{\frac{5}{2}}\,\dfrac{\,K_{\frac{1}{2}} \big(\,\frac{2\pi\,n\,\ell
\,b}{a}\big)} {\sqrt{\ell\,b}} \,.\eeq{finito3} After performing the
sum over $\ell$ the above expression simplifies to \beq P=-
\dfrac{3\,\zeta(4)}{8\pi^2 a^4}- \dfrac{\zeta(4)}{8\pi^2
b^4}-\dfrac{\pi}{2\,b\,a^{3}}\,\sum_{n=1}^{\infty}
n^2\,\ln(1-e^{-2\,\pi\,n\,b\,/a})\,.\eeq{p} The first term
represents the force per unit area between parallel plates i.e. \beq
P_{||}=- \dfrac{3\,\zeta(4)}{8\pi^2 \,a^4}\,.\eeq{parallel} The
pressure $P$ expressed in units of $P_{||}$ reduces to the
expression \beq \dfrac{P}{P_{||}} = 1
+\dfrac{1}{3}\bigg(\dfrac{a}{b}\bigg)^4+\dfrac{120}{\pi}\bigg(\dfrac{a}{b}\bigg)
\,\sum_{n=1}^{\infty}
n^2\,\ln(1-e^{-2\,\pi\,n\,b\,/a})\,.\eeq{ratio} We plot $P/P_{||}$
as a function of $a/b$ in Fig. 1. The Casimir pressure on the strip
is greater than or equal to one and increases as $a/b$ increases,
reaching a value that is $26\%$ higher than the parallel plate case
when $b=a$.

\section{Summary and discussion}

In this paper we obtain two exact $d$-dimensional expressions for
the PMC Casimir piston namely equations \reff{FPMC} and
\reff{FPMCalt}. We showed that the application of these formulas to
2+1 and 3+1 dimensions is in agreement with previous Dirichlet and
PEC piston results. Moreover, as a spin-off, we obtain an
alternative expression for the 3+1 dimensional PEC Casimir piston
which is useful when the plate separation is larger than the
dimension of the plates. We also calculated the Casimir force per
unit area for the special case of an infinite strip, a geometry of
experimental interest. We showed that the Casimir pressure on the
strip is $26\%$ stronger compared to the pressure on parallel plates
when the side $b$ of the strip equals the plate separation $a$.

The important role that Casimir energies can play when extra
dimensions are present has recently been highlighted in
\cite{Greene}. It was argued that in a brane world scenario with
toroidal extra dimensions, Casimir energies under certain conditions
could stabilize the extra dimensions, allow three dimensions to grow
large and provide an effective dark energy in the large dimensions.
Higher-dimensional Casimir formulas derived in previous works were
used and this illustrates the relevance of such results to
investigations in different branches of Physics.

Driven in large part by communication technologies, the last four to
five years have seen a great interest in structures which
approximate PMC's \cite{antenna}. Casimir experiments involving such
structures may therefore be possible in the not too distant future.
In practice, experiments would yield different results between PEC
and PMC pistons because one is comparing metals with finite electric
conductivity to approximate PMC's with finite magnetic conductivity.
In PEC's, we know that finite electric conductivity corrections can
contribute on the order of  10 to 20 \% of the net Casimir force for
parallel plates separated by approximately $1 \mu m$
\cite{Bordagreport}. It would therefore be worthwhile to calculate
the effects of finite magnetic conductivity on PMC Casimir energies
first in a parallel plate scenario and then a piston scenario. This
is work for the future.
\begin{appendix}

\section{Alternative expressions for the $d+1$-dimensional PMC Casimir piston}
\def\theequation{A.\arabic{equation}}
\setcounter{equation}{0}

We can develop an alternative formula for the PMC Casimir force by
simply labeling the $d$ lengths $L_1, L_2,..,L_d$ in region I
differently while keeping the same labeling for region II. This will
not alter the Casimir energy in region I because it is invariant
under permutation of lengths. In our previous derivation leading to
the $F_{\!P\!M\!C\!}$, eq. \reff{FPMC}, we labeled the $d$ lengths
in region I in the following fashion: $L_1=a_1,
L_2=a_2,...,L_{d-1}=a_{d-1}$ and $L_d=a$ where $a$ is the plate
separation. We now label them $L_1=a, L_2=a_1,...,L_{d}=a_{d-1}$.
Note that this is the same labeling we had for region $I\!I$ in our
original derivation except that now $L_1$ is $a$ instead of $s-a$.
This means that our alternative expression for the Casimir energy in
region I, $E_I^{alt}$,  can be obtained from the formula for
$E_{I\!I}$ (\reff{EIIPMC} together with \reff{RIIp}) by replacing
$s-a$ by $a$. This yields \beq
E_{I}^{alt}=\sum_{p=1}^{d-1}\,\,\sum_{q=0}^{d-p-1}
\dfrac{\pi\,(d\!-\!1\!-\!2p\!-\!q)}{2^{d-q+1}}\,\,\xi^{\,d-q-1}_{\,1,k_2,k_3,..,
k_p}\,\dfrac{a\,a_{k_2-1}\ldots a_{k_p-1}} {(a_{d-q-1})^{p+1}}\big(
Q_p + R_{{I}_p}^{alt}\big)\,\eeq{EIalt} where $Q_p$ is given by
\reff{Qp} and $R_{I_p}^{alt}$ is obtained from \reff{RIIp} with
$s\!-\!a$ replaced by $a$ i.e.\beq R_{I_p}^{alt}
=\sum_{n=1}^{\infty}\,\sumprime_{\substack{\ell_i=-\infty\\i=1,\ldots,
p}}^{\infty}\dfrac{2\,\,n^{\frac{p+1}{2}}}{\pi}\,\dfrac{\,K_{\frac{p+1}{2}}
\big(\,2\pi\,n\,\sqrt{(\ell_1\frac{a}{a_{d-q-1}})^2+\cdots+(\ell_p\,\frac{a_{k_p-1}}{a_{d-q-1}})^2}\,\,
\,\big)}{\left[(\ell_1\frac{a}{a_{d-q-1}})^2+\cdots+(\ell_p\frac{a_{k_p-1}}{a_{d-q-1}})^2\right]^{\tfrac{p+1}
{4}}}\,. \eeq{RIpalt2}

The alternative expression for the Casimir force in region I is
\beq\begin{split}
F_{I}^{alt}&=-\dfrac{\partial\,E_{I}^{alt}}{\partial\,a}\\&=
-\sum_{p=1}^{d-1}\,\,\sum_{q=0}^{d-p-1}
\dfrac{\pi\,(d\!-\!1\!-\!2p\!-\!q)}{2^{d-q+1}}\,\,\xi^{\,d-q-1}_{\,1,k_2,k_3,..,
k_p}\,\dfrac{a_{k_2-1}\ldots a_{k_p-1}}
{(a_{d-q-1})^{p+1}}\\&\qquad\qquad\qquad\qquad\qquad\qquad\Big\{
\Gamma(\tfrac{p+2}{2})\,\,\pi^{\frac{-p-4}{2}}\,\,
\zeta(p+2)\,+\dfrac{\partial\,}{\partial\,a}\big(a\,R_{I_p}^{alt}\big)\Big\}\,.\end{split}\eeq{FIalt}
The expression for the Casimir force in region $I\!I$ is the same as
before i.e. $F_{I\!I}$ given by eq.\reff{FII}: \beq\begin{split}
F_{I\!I}&=\sum_{p=1}^{d-1}\,\,\sum_{q=0}^{d-p-1}
\dfrac{\pi\,(d\!-\!1\!-\!2p\!-\!q)}{2^{d-q+1}}\,\,\xi^{\,d-q-1}_{\,1,k_2,k_3,..,
k_p}\,\dfrac{a_{k_2-1}\ldots a_{k_p-1}}
{(a_{d-q-1})^{p+1}}\,\\&\fivequad\fivequad\qquad\Big\{
\Gamma(\tfrac{p+2}{2})\,\,\pi^{\frac{-p-4}{2}}\,\,
\zeta(p+2)\,+R_{{I\!I}_p}(\ell_1\!=\!0)\Big\}\end{split}\eeq{FIIA}
where $R_{{I\!I}_p}(\ell_1\!=\!0)$ is given by \reff{RIIp2}. The
alternative expression for the Casimir force on the piston,
$F_{\!P\!M\!C}^{alt}$ is obtained by adding $F_{I}^{alt}$ and
$F_{I\!I}$. Note that the first term in the curly brackets (the term
with the Riemann zeta function) of $F_{I}^{alt}$ and $F_{I\!I}$ are
identical except that one is the negative of the other. They
therefore cancel out. Note also that the $\ell_1=0$ part of the
second term in the curly brackets of $F_{I}^{alt}$ cancels out with
the second term in $F_{I\!I}$ since
\beq-\dfrac{\partial\,}{\partial\,a}\big\{a\,R_{I_p}^{alt}(\ell_1\!=\!0)\big\}=
-R_{I_p}^{alt}(\ell_1\!=\!0)=-R_{{I\!I}_p}(\ell_1\!=\!0)\,.\eeq{ffr}
The alternative expression for the PMC Casimir force reduces to \beq
F_{\!P\!M\!C}^{alt}=F_{I}^{alt} +F_{I\!I}=
-\sum_{p=1}^{d-1}\,\,\sum_{q=0}^{d-p-1}
\dfrac{\pi\,(d\!-\!1\!-\!2p\!-\!q)}{2^{d-q+1}}\,\,\xi^{\,d-q-1}_{\,1,k_2,k_3,..,
k_p}\,\dfrac{a_{k_2-1}\ldots a_{k_p-1}}
{(a_{d-q-1})^{p+1}}\dfrac{\partial\,}{\partial\,a}\big\{a\,R_{I_p}^{alt}(\ell_1\!\ne\!0)\big\}\eeq{FPMCalt}
where $R_{I_p}^{alt}(\ell_1\!\ne\!0)$ is \reff{RIpalt2} evaluated
without including $\ell_1\!=\!0$ i.e. \beq
R_{I_p}^{alt}(\ell_1\!\ne\!0)
=\sum_{n=1}^{\infty}\,\sum_{\ell_1=1}^{\infty}\sum_{\substack{\ell_i=-\infty\\i=2,\ldots,
p}}^{\infty}\dfrac{4\,\,n^{\frac{p+1}{2}}}{\pi}\,\dfrac{\,K_{\frac{p+1}{2}}
\big(\,2\pi\,n\,\sqrt{(\ell_1\frac{a}{a_{d-q-1}})^2+\cdots+(\ell_p\,\frac{a_{k_p-1}}{a_{d-q-1}})^2}\,\,
\,\big)}{\left[(\ell_1\frac{a}{a_{d-q-1}})^2+\cdots+(\ell_p\frac{a_{k_p-1}}{a_{d-q-1}})^2\right]^{\tfrac{p+1}
{4}}}\,. \eeq{RIpalt} In contrast to \reff{RIpalt2}, there is no
longer a prime on the sum over $\ell_i$ and it starts at $i\!=\!2$.

\end{appendix}
\section*{Acknowledgments}

AE acknowledges support from the Natural Sciences and Engineering
Research Council of Canada (NSERC) and VM acknowledges support from
a CNRS grant ANR-06-NANO-062 and grants RNP 2.1.1.1112, SS
5538.2006.2 and RFBR 07-01-00692-a.

\end{document}